\documentclass[12pt,a4paper]{article}

\usepackage{amsmath}
\usepackage{graphicx}
\usepackage{epstopdf}
\usepackage{color}
\usepackage{a4wide}
\usepackage[numbers,square,sort&compress]{natbib}

\def \Grad{\mbox{Grad\hskip 1pt}}
\def \Curl{\mbox{Curl\hskip 1pt}}

\definecolor{purpleheart}{rgb}{0.41, 0.21, 0.61}

\title{Bending control and instability of functionally\\ graded dielectric elastomers}

\author{
Yipin Su$^{1,2,3}$, Ray W. Ogden$^4$, Michel Destrade$^{1,2}$\\[24pt]
$^{1}$\,School of Mathematics, Statistics and Applied Mathematics \\
NUI Galway, University Road, Galway, Ireland\\[2pt]
$^{2}$\,Department of Engineering Mechanics, Zhejiang University\\
Hangzhou 310027, PR China\\[4pt]
$^{3}$
Sonny Astani Department of Civil and Environmental Engineering\\University of Southern California, Los Angeles, CA 90089, USA\\[4pt]
$^4$\,School of Mathematics and Statistics, University of Glasgow\\
University Place, Glasgow G12 8SQ, UK
\\[12pt]
}

\date{}

\begin{document}

\numberwithin{equation}{section}

\maketitle

\begin{abstract}

A rectangular plate of dielectric elastomer exhibiting gradients of material properties through its thickness will deform \emph{inhomogeneously} when a potential difference is applied to compliant electrodes on its major surfaces, because each plane parallel to the major surfaces will expand or contract to a different extent.  
Here we study the voltage-induced bending response of a functionally graded dielectric plate on the basis of the nonlinear theory of electro-elasticity, when both the elastic shear modulus and the electric permittivity change with the thickness coordinate.
The theory is illustrated for a neo-Hookean electro-elastic energy function with the shear modulus and permittivity varying linearly across the thickness.

We find that in general the bending angle increases with the potential difference provided the Hessian remains positive, but instability can arise as the potential difference increases and the Hessian vanishes.  
We derive the Hessian criterion for predicting the instability, and  show how the material gradients  can be tuned to control the bending shape, and to delay or promote the onset of the instability of the elastomer.

\end{abstract}


\newpage


\section{Introduction}


Dielectric elastomers are soft active materials capable of undergoing large deformations rapidly in response to an applied potential difference (voltage) across their thickness, and have therefore attracted considerable academic and industrial attention in recent years \cite{Pelrine2000, O'Halloran2008, Brochu2010, Alessandri2016, Nketia?Yawson2018}. 
\emph{Shape control}, which is eagerly pursued in dielectric elastomers, has considerable potential for applications to, for example, soft robots, energy harvest systems, actuators and sensors \cite{Li2017, Li2018}.   
Hence, based on the concept of \emph{smart-bending actuation}  \cite{Wang2017, Su2019}, an intelligent self-controlled switch can be designed to keep the working voltage in a desired range: as the voltage rises or falls, the material bends toward or away from a contact point so as to moderate the voltage automatically.

One clever strategy for designing voltage-responsive bending solids is to use dielectric elastomers with physical properties that vary in the thickness direction. When subject to a voltage through the thickness, a non-uniform deformation is generated in the material, which can result in a global bending.   Dielectric-based multilayers composed of pre-stretched dielectric and inactive elastic layers have been proposed to realise shape control by  tuning the applied voltage only, without any mechanical input \cite{Wang2017, Su2019}. However, discontinuous shear stresses may arise at the interface of the layers, which may result in slide, exfoliation or crack formation during the bending deformation \cite{Rocca2010, Morimoto2015, Su2019}.

To overcome this potential problem,  this paper proposes to study functionally graded dielectric elastomers (FGDEs) with material properties varying \emph{continuously} through the thickness. \color{black} 
Experimentally, bulk functionally graded materials (FGMs) can be manufactured by methods such as Powder Metallurgy Technique, Centrifugal Casting and Solid Freeform Technology \cite{Knoppers2005}, etc. 
To produce thin FGMs, techniques such as Physical or Chemical Vapour Deposition (PVD/C-VD), Plasma Spraying and Self-propagating High-temperature Synthesis (SHS) are more appropriate \cite{Mahamood2012}. Spatial inhomogeneous dielectric properties can be introduced by embedding electroactive particles into a polymer FGM matrix unevenly through a well-established additive manufacturing (3D printing) process \cite{Wang2017}. The electric field can be generated by applying a voltage through the flexible electrodes covered on the upper and lower faces of the resulting elastomers.
\color{black}

If the gradients of properties vary in a monotone manner though the plate thickness, we expect the expansion or contraction of each plane in the plate to also vary in the same way; hence, \emph{the plate will bend}.
Here we use the framework of nonlinear electro-elasticity \cite{Dorfmann2006,Dorfmann2014} to investigate the nonlinear bending behaviour and  the stability  of such plates.

For an FGDE plate the required analysis, summarised in Section \ref{sec2}, is  complicated by the inherent inhomogeneity, but we nonetheless manage to derive analytical formulas governing the voltage-induced bending of  a plate made of a neo-Hookean dielectric with linear gradients in its material properties. 
In Section \ref{sec3} we obtain numerical results showing the bent shape of the plate and the associated stress distributions.  The results vary significantly with the selected values of the grading parameters.

For a plate with uniform properties, the onset of \emph{pull-in instability} has been examined extensively, see for example the papers by Zhao and Suo \cite{Zhao2007}, Lu et al. \cite{Lu2012}, Zhao and Wang \cite{Zhao2014}, Zurlo et al. \cite{Zurlo2017}, or Su et al. \cite{Su2018, Su2019b}.  
In general, the pull-in instability leads to a dramatic thinning of the plate, thus triggering a giant deformation in the material.  
Here the corresponding phenomenon would be a sudden increase in the bending angle and, in general, a sudden thinning of the bent plate. However, we find (at least for the geometrical and material parameters chosen as examples) that this type of pull-in instability does not arise, because the maximal configuration of a closed circular ring is reached before the maximum in the voltage-stretch relation.

In Section \ref{sec4}, we examine the stability of the bent configuration on the basis of positive definiteness of the second variation of the free energy, for the considered geometry. 
In particular we derive the associated Hessian criterion.
On this basis we find that the considered bent configurations are either unstable, irrespective of the applied (non-zero) voltage, or stable for a limited range of the applied voltage, depending on the values of the grading parameters.  These results, which we summarise in Section \ref{sec5}, have potential for informing the design of high-performance actuators and sensors.


\section{Bending response of an FGDE plate\label{sec2}}


\subsection{Kinematics and electric field}

\begin{figure}[h!]
\centering
\includegraphics[width=0.7\textwidth]{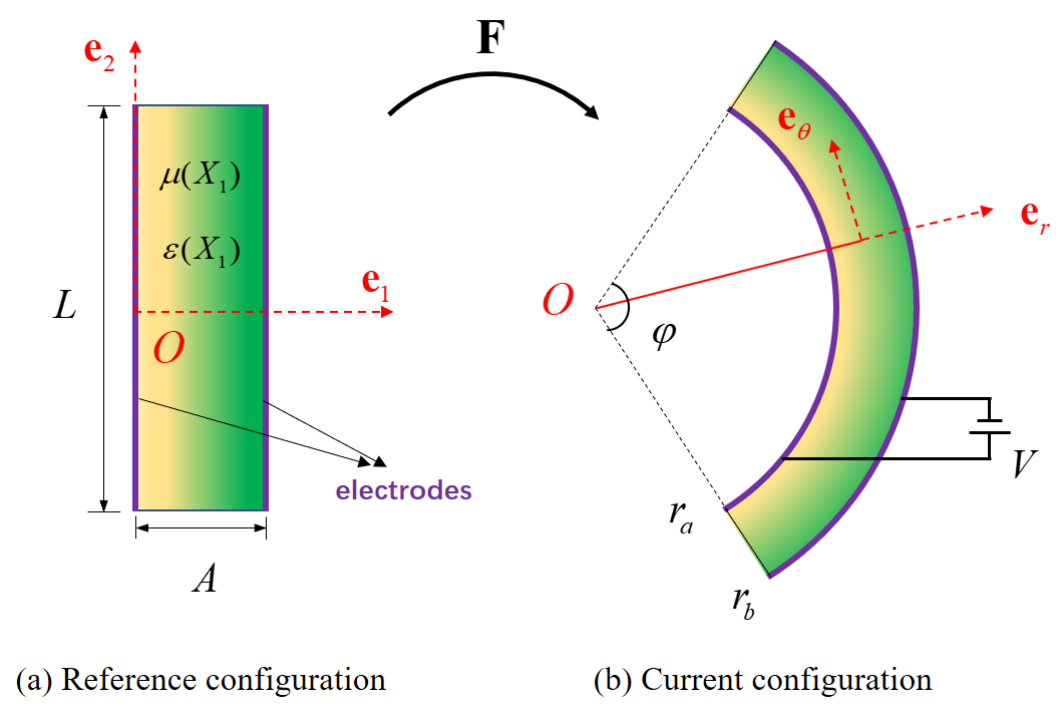}
\caption{
A block of FGDE with continuously varying shear modulus $\mu(X_1)$ and permittivity $\varepsilon(X_1)$ through the thickness, coated with two compliant electrodes. The layer bends when loaded with a voltage $V$ across the electrodes, with the bending angle $\varphi$.
\label{figure1}}
\end{figure}

Consider an FGDE plate of initial length $L$, thickness $A$ and width $H$, the latter being assumed to be much longer than its thickness and length.  The solid occupies the region $0\le X_1\le A$, $-L/2\le X_2\le L/2$, $0 \le X_3 \le H$ in the reference configuration, and the faces $X_1=0,A$ are coated with flexible electrodes as depicted in Figure \ref{figure1}$(a)$.   The material properties are taken to be inhomogeneous with grading dependent on the thickness coordinate $X_1$.  

On application of a potential difference $V$ (voltage) across the electrodes the plate is bent into the shape shown in Figure \ref{figure1}$(b)$, and the resulting deformation is assumed to have a plane strain character in the $(X_1,X_2)$ plane, as in \cite{Wang2017, Su2019}.  In Figure \ref{figure1}$(a)$ we have indicated the dependence on $X_1$ of the shear modulus $\mu$ and permittivity $\varepsilon$, which will be specified in Section \ref{sec2-2}.  The material is assumed to be incompressible.

The  deformation of the plate is described by the equations
\begin{equation} \label{bending}
r=\sqrt{r_a^2 + (r_b^2-r_a^2)\frac{X_1}{A}}, \qquad \theta = \frac{2A}{r_b^2-r_a^2} X_2, \qquad z = X_3,
\end{equation}
as in \cite{Rivlin1949, Su2019c},
where  $(r,\theta, z)$ are the cylindrical polar coordinates in the deformed configuration, and $r_a$, $r_b$ are the radii of the inner and outer bent surfaces, respectively.
The associated deformation gradient, denoted $\mathbf{F}$, has the diagonal form $\mathrm{diag}(\lambda^{-1},\lambda,1)$ with respect to the cylindrical polar axes, where
\begin{equation}
\lambda=\frac{2A r}{r_b^2-r_a^2}\label{lambda-r}
\end{equation}
is the \emph{circumferential stretch}.
It takes the values
\begin{equation}
\lambda_a=\frac{2A r_a}{r_b^2-r_a^2}, \qquad
\lambda_b=\frac{2A r_b}{r_b^2-r_a^2}, \label{lambdaab}
\end{equation}
on the inner and outer bent faces, respectively.

Note that, by taking $X_2=L/2$ and $\theta = \varphi/2$ in \eqref{bending}$_2$,  we obtain an expression for the \emph{bending angle} $\varphi$, namely
\begin{equation}
\varphi=(\lambda_b^2-\lambda_a^2)L/(2A),\label{anglevarphi}
\end{equation}
as given in
\cite{Su2019c}.  Hence $\lambda=\varphi r/L$.  

We shall need $X_1$ in terms of $\lambda$ later.  On use of \eqref{bending}$_1$, \eqref{lambda-r} and \eqref{lambdaab} this is written as
\begin{equation}
X_1=\frac{\lambda^2-\lambda_a^2}{\lambda_b^2-\lambda_a^2}A.\label{X1tolambda}
\end{equation}

On application of the voltage $V$ a radial electric field component, denoted $E$, is generated in the bent configuration of the material, assuming that edge effects can be neglected, and the corresponding electric displacement field component is denoted $D$.  Each of $E$ and $D$ depends on $r$ and is independent of $\theta$ and $z$.  Maxwell's equation $\text{div }\mathbf{D}=0$ then reduces to $\mathrm{d}(rD)/\mathrm{d}r=0$, so that $rD$ is constant.  The corresponding Lagrangian field, given by $\mathbf{D}_\mathrm{L}=\mathbf{F}^{-1}\mathbf{D}$ in general for an incompressible material, reduces to the single component $D_\mathrm{L}=\lambda D$, which by \eqref{lambda-r} is therefore a \emph{constant}.


\subsection{Constitutive law\label{sec2-2}}


For an isotropic electro-elastic material in general we take the energy to be a function of $\mathbf{F}$ and $\mathbf{D}_\mathrm{L}$, but for the considered geometry, deformation and electric field, this reduces to dependence on $\lambda$ and $D_\mathrm{L}$.  We denote the energy function by $\omega^*(\lambda,D_\mathrm{L})$.  

The relevant components of the total Cauchy stress tensor are the radial and circumferential components, denoted $\tau_{rr}$ and $\tau_{\theta\theta}$, respectively.  These satisfy the equilibrium equation
\begin{equation}
\frac{\mathrm{d}}{\mathrm{d}r}(r\tau_{rr})=\tau_{\theta\theta}.\label{equil-r}
\end{equation}
The stress difference $\tau_{\theta\theta}-\tau_{rr}$ and Lagrangian electric field $E_\mathrm{L}=\lambda^{-1}E$ are obtained from the formulas
\begin{equation}
\tau_{\theta\theta}-\tau_{rr}=\lambda\frac{\partial \omega^*}{\partial\lambda},\qquad E_\mathrm{L}=\frac{\partial\omega^*}{\partial D_\mathrm{L}};\label{const-diff}
\end{equation}
see, for example,  \cite{Dorfmann2006,Dorfmann2014}.

Since, from \eqref{lambda-r}, $\lambda$ is proportional to $r$, while $D_\mathrm{L}$ is constant, the combination of \eqref{equil-r} and \eqref{const-diff}$_1$ followed by integration, leads to
\begin{equation}\label{normal-stress}
\tau_{rr}=\omega^*+K, \qquad \tau_{\theta\theta}=\lambda\frac{\partial \omega^*}{\partial \lambda}+\omega^*+K,
\end{equation}
as derived in  \cite{Su2019c},
where $K$ is a constant to be determined from the boundary conditions, in a similar way to the situation in the purely elastic case \cite{Rivlin1949}.

Here we assume that the inner and outer surfaces of the bent sector at $r_a$ and $r_b$ are free of mechanical traction, so that
\begin{equation}
\tau_{rr}(r_a)= 0, \qquad \tau_{rr}(r_b)=0,
\label{traction-free}
\end{equation}
there being no Maxwell stress on these surfaces.

For definiteness, we consider an energy function which has the form
\begin{equation}
\omega^*(\lambda,D_\mathrm{L})=W(\lambda)+\frac{D_\mathrm{L}^2}{2\varepsilon\lambda^2},\label{model-W-D}
\end{equation}
where $W$ is derived from any isotropic purely elastic strain-energy function and we recall that $D=\lambda^{-1} D_\mathrm{L}$.

The boundary conditions \eqref{traction-free} provide two expressions for $K$, 
\begin{equation}
-K=W(\lambda_a)+\frac{D_\mathrm{L}^2}{2\varepsilon_a\lambda_a^2}=W(\lambda_b)+\frac{D_\mathrm{L}^2}{2\varepsilon_b\lambda_b^2},
\end{equation}
and hence an expression for $D_\mathrm{L}^2$, namely
\begin{equation}
D_\mathrm{L}^2=2\varepsilon_a\varepsilon_b\lambda_a^2\lambda_b^2 \frac{W(\lambda_b)-W(\lambda_a)}{\varepsilon_b\lambda_b^2-\varepsilon_a\lambda_a^2},\label{DLsquared}
\end{equation}
where $\varepsilon_a$ and $\varepsilon_b$ are the values of $\varepsilon$ at $r=r_a$ and $r=r_b$, respectively, i.e. for $X_1=0, A$.  The form of the function $\varepsilon(X_1)$ will be exemplified below.
It follows that
\begin{equation}
K=-\frac{\varepsilon_b\lambda_b^2W(\lambda_b)-\varepsilon_a\lambda_a^2W(\lambda_a)}{\varepsilon_b\lambda_b^2-\varepsilon_a\lambda_a^2}.\label{K}
\end{equation}

Let $\phi(X_1)$ denote the electrostatic potential through the thickness.  Then, from Maxwell's equation $\Curl\mathbf{E}_\mathrm{L}=\mathbf{0}$ we have $\mathbf{E}_\mathrm{L}=-\Grad\phi$, which specializes here to $E_\mathrm{L}=-\mathrm{d}\phi/\mathrm{d}X_1$.  The potential difference $\phi(0)-\phi(X_1)$ is the voltage  $V$, which, on use of \eqref{const-diff}$_2$ is given by
\begin{equation}
V=\int_0^AE_\mathrm{L}\mathrm{d}X_1= \dfrac{2A}{\lambda_b^2-\lambda_a^2}\int_{\lambda_a}^{\lambda_b}\lambda\dfrac{\partial \omega^*}{\partial D_\mathrm{L}}\mathrm{d}\lambda,
\end{equation}
the latter change of variable making use of \eqref{X1tolambda}.
For the model \eqref{model-W-D}, this yields
\begin{equation}
V=\frac{2AD_\mathrm{L}}{\lambda_b^2-\lambda_a^2}\int_{\lambda_a}^{\lambda_b}\frac{\mathrm{d}\lambda}{\lambda\varepsilon}\, ,\label{Vfinal}
\end{equation}
and we recall that $\varepsilon$ depends on $X_1$ and hence on $\lambda$.

To complete the formulation of the problem, we first note that the resultant force on the lateral end faces at $\theta=\pm \varphi/2$ vanishes since, with the help of the boundary conditions \eqref{traction-free} and equation \eqref{equil-r}, it follows that
\begin{equation}
\int_{\lambda_a}^{\lambda_b}\tau_{\theta\theta}\mathrm{d}r=0
\end{equation}
is automatically satisfied  \cite{Su2019c}.  We then assume that there is no resultant moment on these faces so that \emph{the block is bent by the application of a voltage alone}.
This requires
\begin{equation}\label{integral_circumferential}
\int_{r_a}^{r_b}r\tau_{\theta\theta}\mathrm{d}r=0,\quad\mbox{equivalently}\quad \int_{\lambda_a}^{\lambda_b}\lambda\tau_{\theta\theta}\mathrm{d}\lambda=0,
\end{equation}
which, from \eqref{normal-stress}$_2$, yields
\begin{equation}\label{moment}
\int_{\lambda_a}^{\lambda_b}\lambda\left(\lambda\frac{\partial \omega^*}{\partial \lambda}+\omega^*+K\right)\text d\lambda=0,
\end{equation}
and for the model \eqref{model-W-D},
\begin{equation}
\int_{\lambda_a}^{\lambda_b}\lambda(\lambda W_\lambda +W)\mathrm{d}\lambda-\frac{1}{2}D_\mathrm{L}^2\int_{\lambda_a}^{\lambda_b}\left(
\frac{1}{\lambda\varepsilon}+\frac{\varepsilon_\lambda}{\varepsilon^2}\right)
\mathrm{d}\lambda
+\frac{1}{2}K(\lambda_b^2-\lambda_a^2)=0,\label{abc-connection}
\end{equation}
where the subscript $\lambda$ signifies the derivative with respect to $\lambda$.

Equations  \eqref{DLsquared} and \eqref{K}, respectively,  give $D_\mathrm{L}$ and $K$ in terms of $\lambda_a$ and $\lambda_b$.  Equation \eqref{Vfinal} then determines $V$ also in terms of
$\lambda_a$ and $\lambda_b$, which we write in the form
\begin{equation}
V=f(\lambda_a,\lambda_b).\label{V-f}
\end{equation}
Once the form of $W$ is prescribed, equation \eqref{abc-connection}, after substituting for $D_\mathrm{L}$ and $K$, yields an implicit connection between $\lambda_a$ and $\lambda_b$, which we write as
\begin{equation}
g(\lambda_a,\lambda_b)=0.\label{g-0}
\end{equation}
Then, in principle, $V$ can be determined in terms of $\lambda_a$ (or $\lambda_b$).

For further specialization we now take $W$ to be the \emph{neo-Hookean strain-energy function}, which, for the considered plane deformation, has the form
\begin{equation}
W(\lambda)=\frac{\mu}{2}(\lambda^2+\lambda^{-2}-2),\label{neo-Hookean}
\end{equation}
where the shear modulus $\mu$ is functionally graded through the thickness, i.e. is a function of $X_1$.

Then the formulas \eqref{DLsquared} and \eqref{K} specialize to
\begin{equation}
D_\mathrm{L}^2=\varepsilon_a\varepsilon_b \frac{\mu_b\lambda_a^2(\lambda_b^2-1)^2-\mu_a\lambda_b^2(\lambda_a^2-1)^2}{\varepsilon_b\lambda_b^2-\varepsilon_a\lambda_a^2},\label{DLsquared2}
\end{equation}
\begin{equation}
K=-\frac{1}{2}\frac{\mu_b\varepsilon_b(\lambda_b^2-1)^2-\mu_a\varepsilon_a(\lambda_a^2-1)^2}{\varepsilon_b\lambda_b^2-\varepsilon_a\lambda_a^2},\label{K+}
\end{equation}
where $\mu_a$ and $\mu_b$ are the values of $\mu(X_1)$ at $X_1=0,A$ ($r=r_a,r_b$),
and \eqref{abc-connection} becomes
\begin{equation}
\int_{\lambda_a}^{\lambda_b}\mu(3\lambda^3-\lambda^{-1}-2\lambda)
\mathrm{d}\lambda-D_\mathrm{L}^2\int_{\lambda_a}^{\lambda_b}\left(
\frac{1}{\lambda\varepsilon}+\frac{\varepsilon_\lambda}{\varepsilon^2}\right)
\mathrm{d}\lambda
+K(\lambda_b^2-\lambda_a^2)=0.\label{abc-connection2}
\end{equation}

To take this further we need to specify how $\mu$ and $\varepsilon$ depend on $X_1$, and through the connection \eqref{X1tolambda}, on $\lambda$.  We assume a simple \emph{linear dependence} of $\mu$ and $\varepsilon$ on $X_1$, as in \cite{Wu2017, Bayat2019}, here specified by
\begin{equation}
\mu(X_1)=\mu_a\left(1-\alpha_\mu X_1/A\right), \qquad \varepsilon(X_1)=\varepsilon_a\left(1+\alpha_\varepsilon X_1/A\right),\label{linear-law}
\end{equation}
with $\mu_b=\mu_a(1-\alpha_\mu)$ and $\varepsilon_b=\varepsilon_a(1+\alpha_\varepsilon)$, where $\alpha_\mu$ and $\alpha_\varepsilon$ are two dimensionless constants characterizing the functionally graded properties of the shear modulus and permittivity, respectively, of the material.  The expressions for $D_\mathrm{L}^2$ and $K$ in \eqref{DLsquared2} and \eqref{K+} are then modified accordingly.  The parameters $\alpha_\mu$ and $\alpha_\varepsilon$ are subject to the restrictions $\alpha_\mu<1$ and  $\alpha_\varepsilon>-1$ since both $\mu$ and $\varepsilon$ are positive.   Additionally, since the permittivity of all materials is greater than the vacuum permittivity $\varepsilon_0$, we must have $\varepsilon_a(1+\alpha_\varepsilon)>\varepsilon_0$ for each $\alpha_\varepsilon>-1$.  Note that negative values of $\alpha_\mu$ and $\alpha_\varepsilon$ reverse the direction of bending, as is illustrated in Section \ref{sec4}.

The integral in \eqref{Vfinal} can now be evaluated to give
\begin{equation}
V=\frac{AD_\mathrm{L}}{\varepsilon_a[(1+\alpha_\varepsilon)\lambda_a^2-\lambda_b^2]}\log\left[(1+\alpha_\varepsilon)\frac{\lambda_a^2}{\lambda_b^2}\right].\label{Vlast}
\end{equation}

Equation \eqref{abc-connection2} can also be evaluated to give
\begin{multline}
\mu_a\frac{[\lambda_b^2-(1-\alpha_\mu)\lambda_a^2]}{\lambda_b^2-\lambda_a^2}\left[\frac{3}{4}(\lambda_b^4-\lambda_a^4)-\log\left(\frac{\lambda_b}{\lambda_a}\right)-(\lambda_b^2-\lambda_a^2)\right]
\\[2pt]
-\frac{1}{2}\mu_a\alpha_\mu\left[\lambda_b^4+\lambda_a^2\lambda_b^2+\lambda_a^4-(\lambda_b^2+\lambda_a^2)-1\right]+K(\lambda_b^2-\lambda_a^2)
\\[2pt]
-\frac{1}{2}D_\mathrm{L}^2\frac{\lambda_b^2-\lambda_a^2}{\varepsilon_a[\lambda_a^2(1+\alpha_\varepsilon)-\lambda_b^2]}\log\left[(1+\alpha_\varepsilon)\frac{\lambda_a^2}{\lambda_b^2}\right]-D_\mathrm{L}^2\frac{\alpha_\varepsilon}{\varepsilon_a(1+\alpha_\varepsilon)}
=0. \label{g-specific}
\end{multline}

For a given applied voltage $V$, after $D_\mathrm{L}$ is substituted into \eqref{Vlast} and $D_\mathrm{L}^2$ and $K$ into \eqref{g-specific} from \eqref{DLsquared2} and \eqref{K+}, the resulting equations yield specific forms of the functions $f$ and $g$ in \eqref{V-f} and \eqref{g-0}, which can now in principle be solved  simultaneously for $\lambda_a$ and $\lambda_b$.  The dependence of the  bending angle $\varphi$, given by \eqref{anglevarphi}, on the applied voltage can then be determined.
The corresponding radial and circumferential stress components $\tau_{rr}$ and $\tau_{\theta\theta}$ can also be determined, from \eqref{normal-stress} with the specializations \eqref{model-W-D} and \eqref{neo-Hookean}.  Numerical results for specific values of  $\alpha_\mu$ and $\alpha_\varepsilon$  are provided in the following section.


\section{Numerical results\label{sec3}}


For numerical purposes we work in terms of the dimensionless quantities:
\begin{equation}
\bar{K} = K/\mu_a,\qquad  \bar{V}= \sqrt{\varepsilon_a/\mu_a}V/A,
\end{equation}
and equations \eqref{V-f} and \eqref{g-0} take the dimensionless forms
\begin{equation}
 \bar{V}=\bar{f}(\lambda_a,\lambda_b),\qquad \bar{g}(\lambda_a,\lambda_b)=0.
 \end{equation}
  These are the dimensionless forms of \eqref{Vlast} and \eqref{g-specific} after substitution for $D_\mathrm{L}$ and $K$, where
 \begin{equation}
 \bar{f}= \sqrt{\varepsilon_a/\mu_a} f/A,\qquad \bar{g}(\lambda_a,\lambda_b)=g(\lambda_a,\lambda_b)/\mu_a.
  \end{equation}
 Thus, $\bar{f}$ and $\bar{g}$ depend only on $\lambda_a,\lambda_b$ and the parameters $\alpha_\mu$ and $\alpha_\varepsilon$, i.e. they are independent of $\mu_a$ and $\varepsilon_a$.

\begin{figure}[!b]
\centering
\includegraphics[width=2.5in]{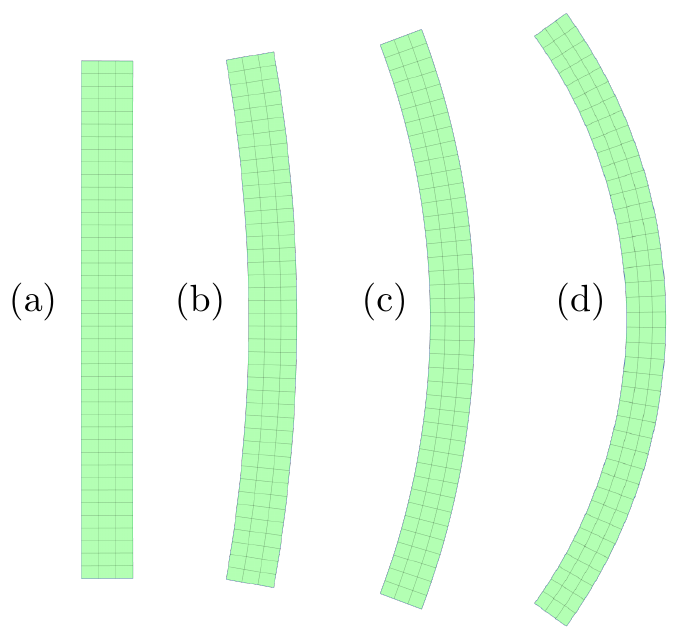}
\caption{(a) Undeformed rectangular plate cross section with aspect ratio $L/A=10$. (b)--(d) plate bent by application of a dimensionless voltage $\bar{V}=0.2$ with the following values of ($\alpha_\varepsilon,\alpha_\mu)$: (b) $(0.1,0.3)$; (c) $(0.4,0.3)$: (d) $(0.7,0.3)$.\label{figure3}}
\end{figure}

\begin{figure}[!h]
\centering
\includegraphics[width=6.4in]{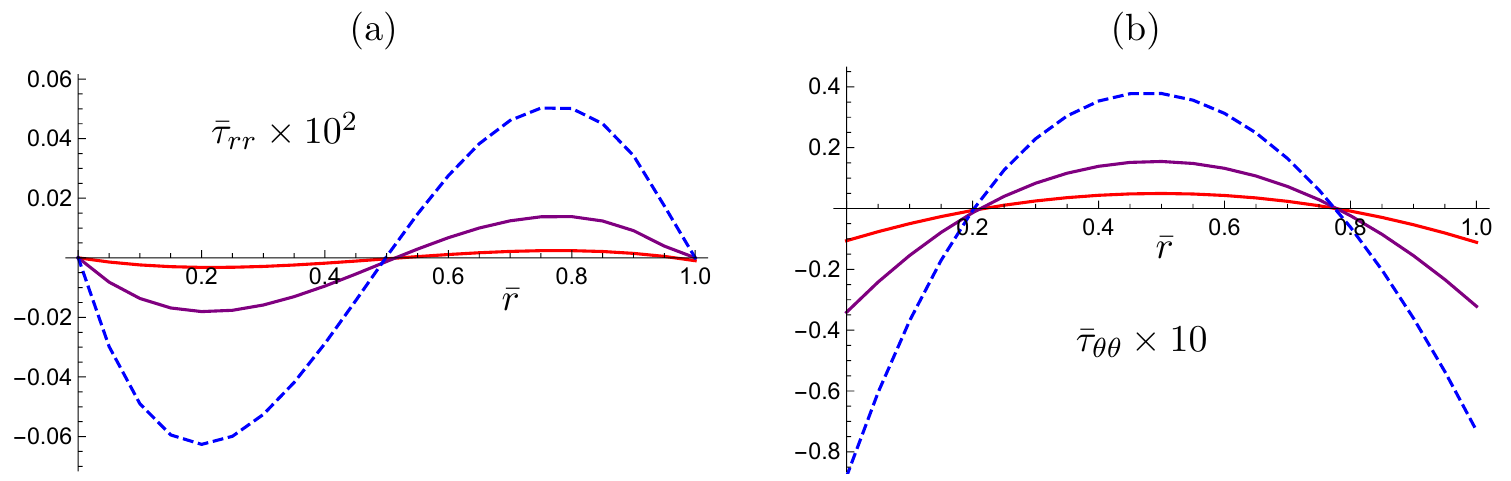}
\caption{Plots of (a) the radial stress $\tau_{rr}$ and (b) the circumferential stress $\tau_{\theta\theta}$ in scaled dimensionless form versus the normalized radius $\bar{r}$ for the dimensionless voltage $\bar{V}=0.2$ and the following values of ($\alpha_\varepsilon,\alpha_\mu)$: red $(0.1,0.3)$; purple $(0.4,0.3)$; blue dashed $(0.7,0.3)$.\label{figure33}}
\end{figure}

For an initially rectangular FGDE plate with aspect ratio $L/A=10$, subject to a (dimensionless) voltage $\bar{V}=0.2$, Figure \ref{figure3} depicts the resulting bent shape of the plate for a fixed value of the elastic parameter $\alpha_\mu=0.3$ and three values of  electric parameter $\alpha_\varepsilon$.  Again for $\alpha_\mu=0.3$ and three values of $\alpha_\varepsilon$, in Figure \ref{figure33} the distributions of the radial and circumferential stress components through the plate are shown in dimensionless forms $\bar{\tau}_{rr}=\tau_{rr}/\mu_a$, $\bar{\tau}_{\theta\theta}=\tau_{\theta\theta}/\mu_a$ versus the radial coordinate in the form $\bar{r}=(r-r_a)/(r_b-r_a)$.  

As expected intuitively, we find that increasing $\alpha_\varepsilon$ makes the plate more susceptible to bending.  The stress components vary continuously through the thickness, with the circumferential stress two orders of magnitude larger than that of the radial stress.  Interestingly, at least for examples considered here, the bent sector has two neutral surfaces (where $\tau_{\theta\theta}=0$).
Additionally we note that while on both its inner and outer faces the circumferential stress is \emph{compressive}, these faces are nonetheless in \emph{extension} as the circumferential stretches are greater than $1$ there.
Similar phenomena were observed for the bending of purely elastic layered plates \cite{Rocca2010} and swelling functionally graded  hydrogel plates \cite{Bayat2019}.

\begin{figure}[!h]
\centering
\includegraphics[width=4in]{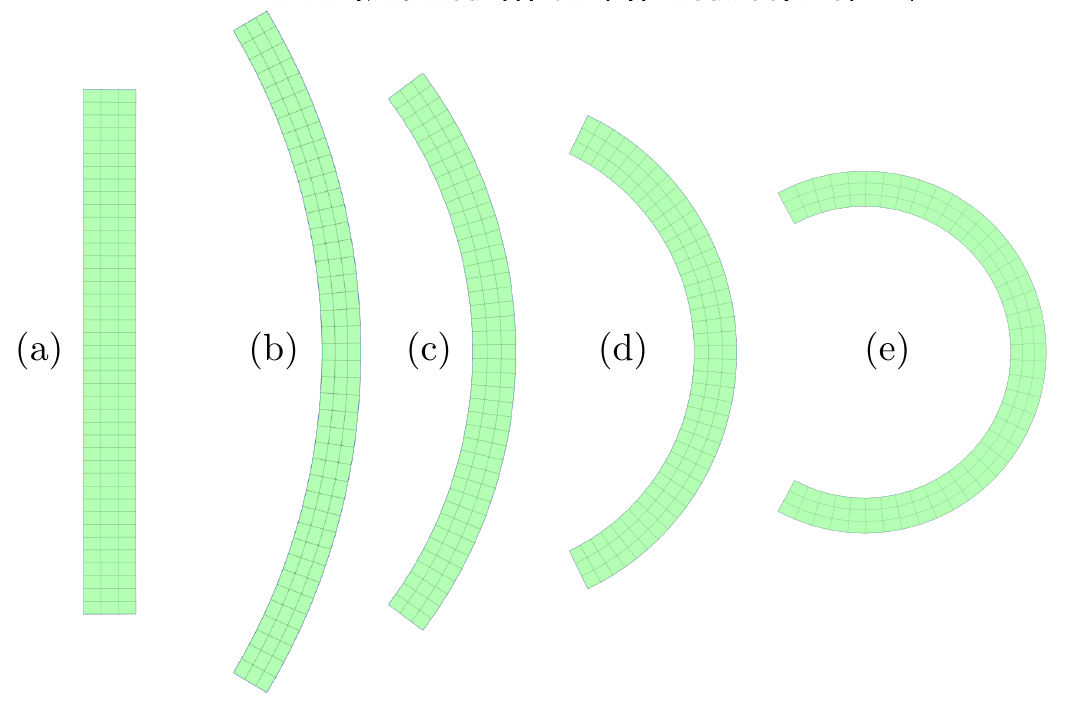}
\caption{(a) Undeformed rectangular plate cross section with aspect ratio $L/A=10$. (b)--(d) plate bent by application of a dimensionless voltage $\bar{V}=0.25$ with the following values of $(\alpha_\varepsilon,\alpha_\mu)$: (b) $(0.4,0.2)$; (c) $(0.4,0.4)$; (d) $(0.4,0.6)$,  and (e) with $\bar{V}=0.4$ and $\alpha_\varepsilon= 0.4, \alpha_\mu=0.4$.
\label{figure44}}
\end{figure}

\begin{figure}[!h]
\centering
\includegraphics[width=6.4in]{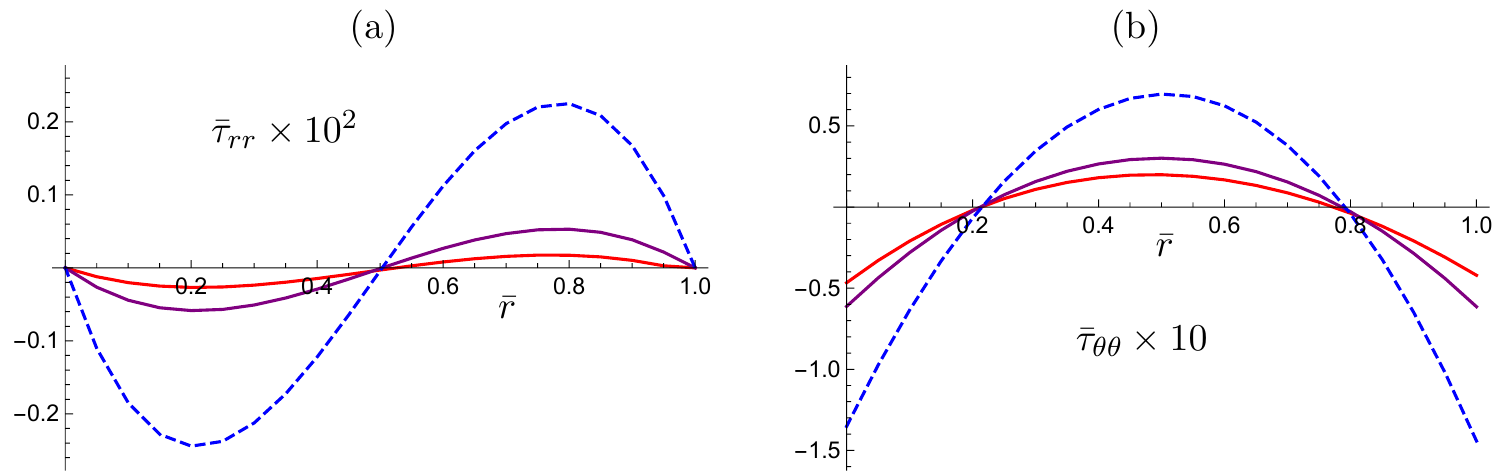}
\caption{Plots of (a) the radial stress $\tau_{rr}$ and (b) the circumferential stress $\tau_{\theta\theta}$ in scaled dimensionless form versus the normalized radius $\bar{r}$ for the dimensionless voltage $\bar{V}=0.25$ and the following values of ($\alpha_\varepsilon,\alpha_\mu)$: red $(0.4,0.2)$; purple $(0.4,0.4)$; blue dashed $(0.4,0.6)$.\label{figure45}}
\end{figure}

For a fixed value of the electric parameter $\alpha_\varepsilon=0.4$ and three values of the elastic parameter  $\alpha_\mu$, results analogous to those in Figures \ref{figure3} and \ref{figure33} are shown in Figures \ref{figure44} and \ref{figure45}, respectively (now with $\bar{V}=0.25$).    In general, whether bending is enhanced or retarded by a change in $\alpha_\mu$ depends on the value of $\alpha_\varepsilon$.

\begin{figure}[!h]
\centering
\includegraphics[width=6.4in]{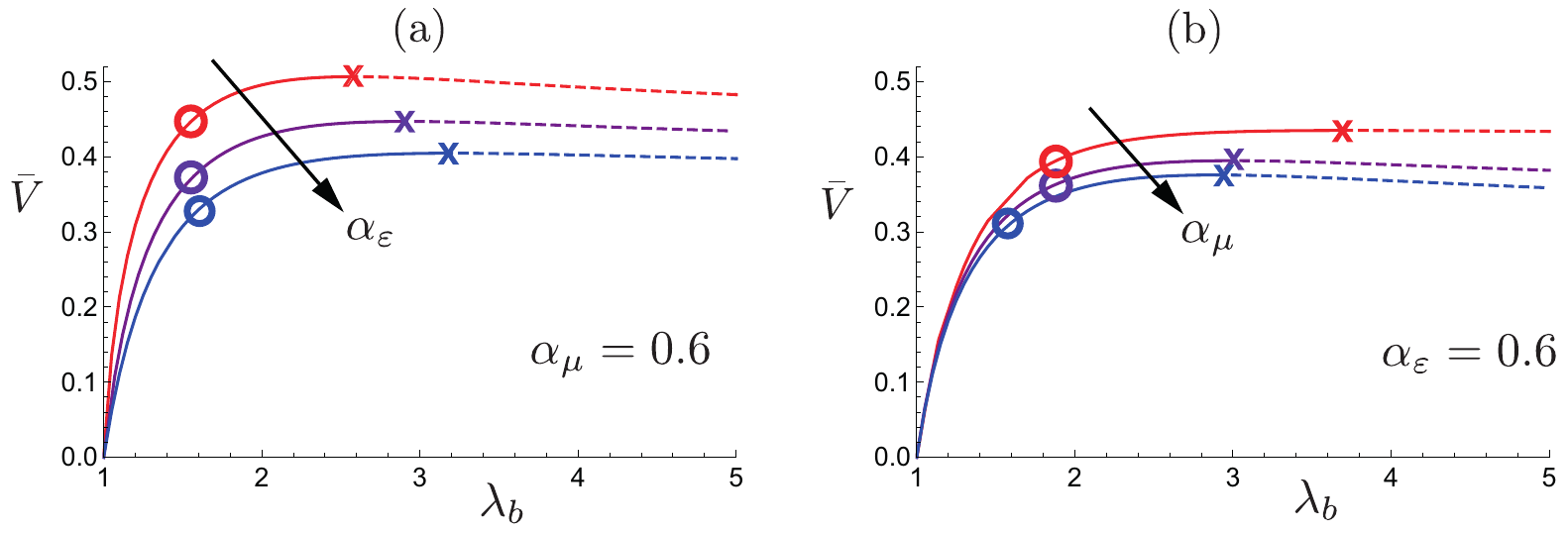}
\caption{Plots of the voltage $\bar{V} $ versus the outer side circumferential stretch $\lambda_b$. (a) Plates with  elastic parameter $\alpha_\mu=0.6$ and increasing $\alpha_\varepsilon=0.1,0.4,0.7$, red, purple and blue, respectively; (b) plates with $\alpha_\varepsilon=0.6$ and increasing elastic parameter $\alpha_\mu=0.4, 0.5,0.8$,  red, purple and blue, respectively.
The circles correspond to points at which a complete circular ring is formed. The maxima of  $\bar{V}$, marked with crosses, are beyond these points and are thus never attained.
\label{figure6}}
\end{figure}

As is clear from the larger value of $\bar{V}$ in Figure \ref{figure44}, as the magnitude of the voltage increases the circular sector becomes more and more bent.  As the voltage increases further the sector will eventually form a \emph{complete circular ring}, which, for the values $\alpha_\varepsilon= 0.4, \alpha_\mu=0.4$, occurs when $\bar{V}$ reaches the approximate value $0.4385$.  At this point the voltage, however, has not reached its maximum value.  Figure \ref{figure6} shows plots of $\bar{V}$ versus $\lambda_b$ for different values of $\alpha_\varepsilon$ and  $\alpha_\mu$, with the  points at which a complete circular ring is formed identified by circles, while the crosses mark the (non-accessible) points at which $\bar{V}$ reaches a maximum.

As noted above, it is also seen from Figures \ref{figure33} and \ref{figure45}  that the inner and outer faces of the bent sector are subject to compressive stress ($\tau_{\theta \theta}<0$) during the bending deformation for the lower values $\bar{V}=0.2, 0.25$ before the complete ring forms.

It is of interest to examine at what point the voltage does indeed reach its maximum, and for this purpose $V$ is considered to be a function of $\lambda_b$ as the independent variable after elimination of $K$ and $D_\mathrm{L}$.  Then,
\begin{equation}
\frac{\mathrm{d}V}{\mathrm{d}\lambda_b}=\left(\frac{\partial f}{\partial \lambda_b}\frac{\partial g}{\partial \lambda_a}-\frac{\partial f}{\partial \lambda_a}\frac{\partial g}{\partial \lambda_b}\right)\Big/\frac{\partial g}{\partial \lambda_a},
\end{equation}
so that a maximum arises when
\begin{equation}
\det\left[\begin{matrix}
 \dfrac{\partial f}{\partial\lambda_a} &  \dfrac{\partial f}{\partial\lambda_b} \\[6pt]
  \dfrac{\partial g}{\partial\lambda_a} &  \dfrac{\partial g}{\partial\lambda_b}\
\end{matrix}\right]=0
\end{equation}
has nontrivial solutions.

However, because the maximum of $V$ does not occur before the plate is bent into a full circle, it is necessary to examine when instability might arise, which, for example, might occur for given values of the parameters $\alpha_\mu$ and $\alpha_\varepsilon$ as the voltage increases.  This is  examined in the next section, based on an energy approach.  

As the maximum of $V$ does not occur in the relevant range of deformations we emphasise that, in contrast to what happens when a homogeneous dielectric plate, equi-biaxially deformed under an applied voltage, \emph{the phenomenon of pull-in instability does not appear for a functionally graded plate} under plane strain.
We recall that in the homogeneous plane strain expansion of \color{black} a  neo-Hookean dielectric plate, \color{black} there is also no pull-in instability, because then $\bar V = \sqrt{1-\lambda^{-4}}$, which  is a monotone function of the stretch \color{black} \cite{Su2019b}\color{black}.
Here our bending deformation \eqref{bending} is also a plane strain deformation, but the $\bar V - \lambda_b$ relationship is not monotone. 

\begin{figure}[!h]
\centering
\includegraphics[width=2.5in]{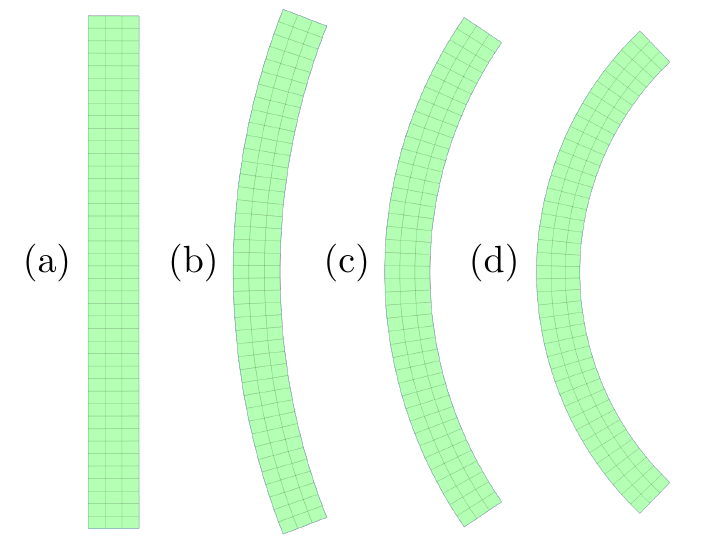}
\caption{(a) Undeformed rectangular plate cross section with aspect ratio $L/A=10$ and $\hat{V}=0.3$, with parameters $(\alpha_\varepsilon,\alpha_\mu)$: (b) $(-0.2,0.3)$; (c) $(-0.4,0.3)$; (d) $(-0.6,0.3)$.
\label{rev2}}
\end{figure}

Because of the dependence of $\mu$ and $\varepsilon$ on $X_1$ there is a lack of symmetry with respect to $X_1=0$ and $X_1=A$.  Thus, bearing in mind the limitations on the values of $\alpha_\mu$ and $\alpha_\varepsilon$ mentioned in Section \ref{sec2-2}, it is possible to consider negative values of  $\alpha_\mu$ and/or $\alpha_\varepsilon$, which leads to a bending response opposite to that shown in Figures \ref{figure3} and \ref{figure44}.  This is illustrated in Figure \ref{rev2} for the fixed value $\alpha_\mu=0.3$ and three negative values of $\alpha_\varepsilon$ with $\bar{V}=0.3$.

\begin{figure}[!h]
\centering
\includegraphics[width=6.4in]{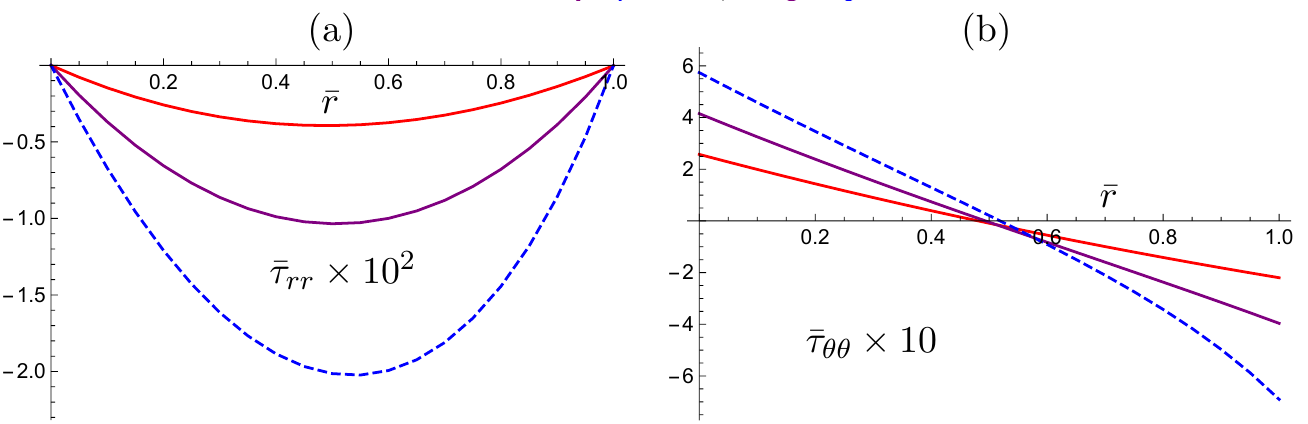}
\caption{Plots of (a) the radial stress $\tau_{rr}$ and (b) the circumferential stress $\tau_{\theta\theta}$ in scaled dimensionless form versus the normalized radius $\bar{r}$ for the dimensionless voltage $\bar{V}=0.3$ and the following values of ($\alpha_\varepsilon,\alpha_\mu)$: red $(-0.2,0.3)$; purple $(-0.4,0.3)$; blue dashed $(-0.6,0.3)$.
\label{revstresses}}
\end{figure}

Figure \ref{revstresses} illustrates the corresponding radial and circumferential stress distributions, which are quite different from those for the situation with both $\alpha_\mu$ and $\alpha_\varepsilon$ positive.  In particular, $\tau_{rr}$ is negative while there is only one neutral surface $\tau_{\theta\theta}=0$. \color{black} Correspondingly, the circumferential stress $\bar \tau_{\theta\theta}$ is monotonic, but it has both positive and negative values and the integral in Eq. \eqref{integral_circumferential} vanishes. \color{black}  The situation is very similar for a fixed negative value of $\alpha_\varepsilon$ with different values of $\alpha_\mu$ or for a fixed pair of values of $\alpha_\varepsilon$ and $\alpha_\mu$ with different voltages.

\begin{figure}[!h]
\centering
\includegraphics[width=4in]{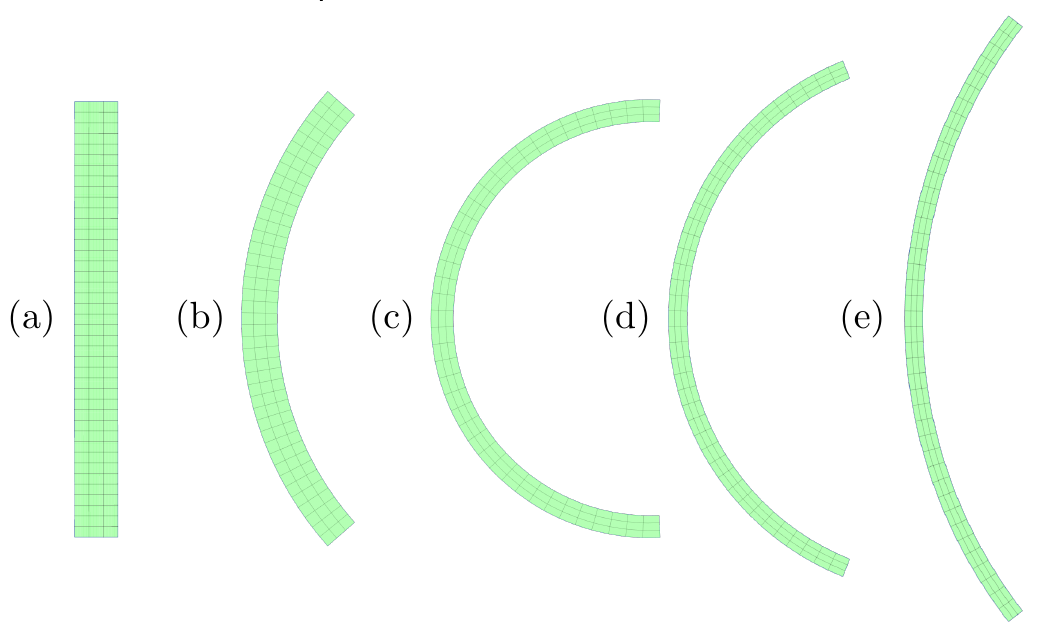}
\caption{(a) Undeformed rectangular plate cross section with aspect ratio $L/A=10$ and parameters $\alpha_\varepsilon=-0.2$, $\alpha_\mu=0.3$ and the following values of $\hat{V}$: (b) $0.5$; (c) $0.87$; (d) $0.89$; (e) $0.9$.\label{minus2plus3}}
\end{figure}

In Figure \ref{minus2plus3}, for fixed parameter values $\alpha_\varepsilon=-0.2,\alpha_\mu=0.3$, the deformation of a plate with increasing voltage is depicted.  Initially the plate bends to the right and then thins significantly as it bends further, but then as the voltage increases more the bending is reversed and further thinning occurs.  No maximum of $\bar{V}$ occurs;
the voltage increases monotonically with increasing values of $\lambda_a$ and $\lambda_b$, which are both larger than $1$.  

\begin{figure}[h!]
\centering
\includegraphics[width=0.5\textwidth]{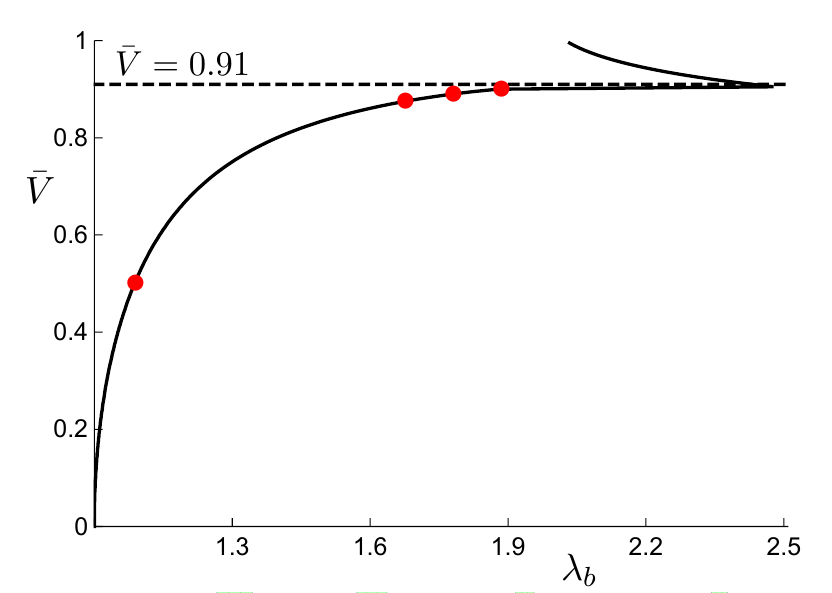}
\caption{Plot of $\bar{V}$ versus $\lambda_b$ for the parameter values $\alpha_\varepsilon=-0.2, \alpha_\mu=0.3$ and $L/A=10$.  The red bullet points are associated with the values $0.5,0.87,0.89,0.9$ of $\bar{V}$ in Figure \ref{minus2plus3}, (b)--(e), respectively.
\label{Vlambda23}}
\end{figure}

A plot of $\bar{V}$ versus $\lambda_b$ is shown in Figure \ref{Vlambda23}.  When $\bar{V}$ reaches a value slightly above $0.903$ the plate becomes straight and thereafter, as  $\bar{V}$ increases further the plate curvature is reversed with $\lambda_b$ decreasing, but then starts to increase again as the reversed curvature increases, as $\bar{V}$ continues to increase.  The latter effect is not shown in Figure \ref{Vlambda23} since the larger values of $\bar{V}$ are impractical.  The red bullet points identify the values $0.5,0.87,0.89,0.9$ of $\bar{V}$ on the curve.

\begin{figure}[!t]
\centering
\includegraphics[width=0.7\textwidth]{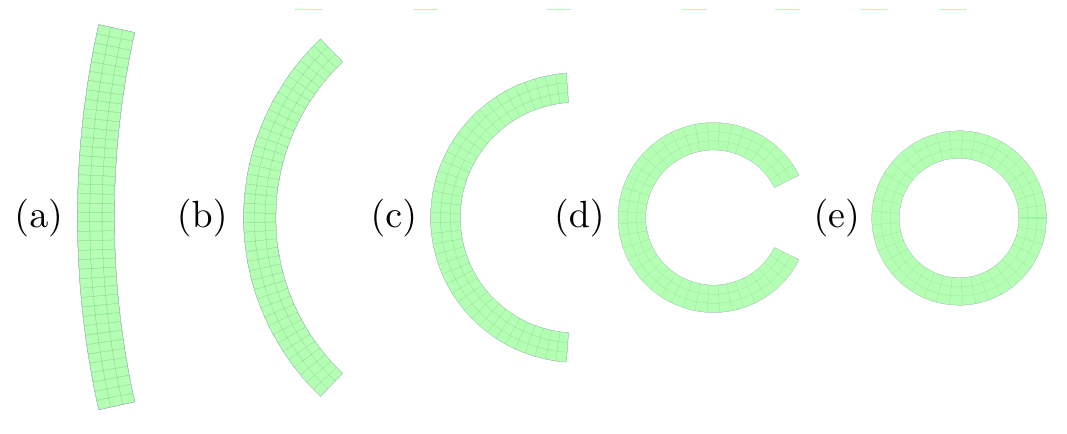}
\caption{Undeformed rectangular plate cross section with aspect ratio $L/A=10$ and parameters $\alpha_\varepsilon=-0.6,\alpha_\mu=0.3$ and the following values of $\hat{V}$: (a) $0.1$; (b) $0.3$; (c) $0.45$; (d) $0.6$; (e) $0.639$.\label{minus6plus3}}
\end{figure}

In Figure \ref{minus6plus3}, for fixed values $\alpha_\varepsilon=-0.6,\alpha_\mu=0.3$, by contrast, the plate bends increasingly to the right without significant thinning until the voltage reaches the approximate value $0.639$, at which point a complete circular ring is formed, while the voltage increases monotonically with increasing values of $\lambda_a$ and $\lambda_b$, which are both larger than $1$. 
Note that the shapes in Figures \ref{figure3}, \ref{figure44}, \ref{rev2}, \ref{minus2plus3}, \ref{minus6plus3} are not drawn exactly to the correct scale, as the areas should all be the same.

\begin{figure}[!h]
\centering
\includegraphics[width=3in]{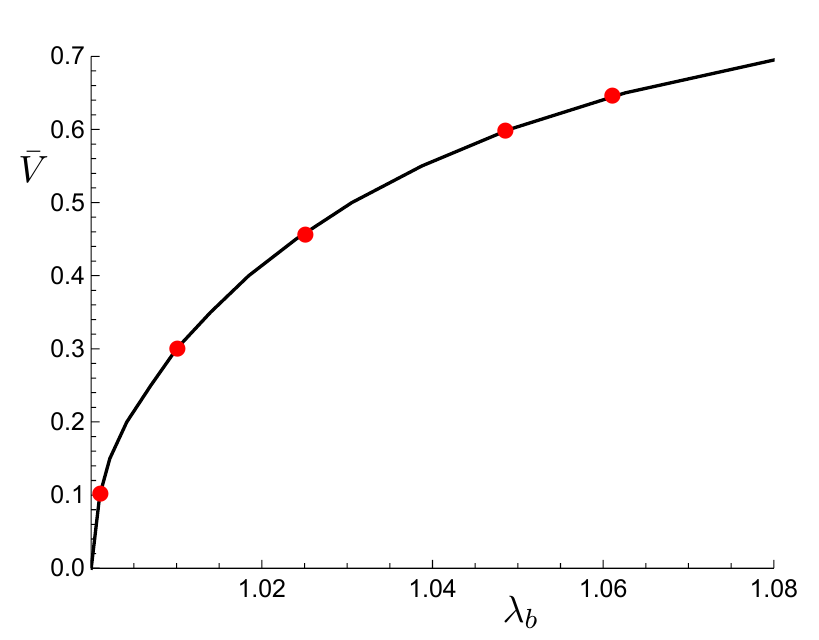}
\caption{Plot of $\bar{V}$ versus $\lambda_b$ for the parameters  $\alpha_\varepsilon=-0.6, \alpha_\mu=0.3$ and $L/A=10$.  The red bullet points are associated with the values $0.1,0.3,0.45,0.6,0.639$ of $\bar{V}$ in Figure \ref{minus6plus3}, (a)--(e), respectively.
\label{Vlambda63}}
\end{figure}

Figure \ref{Vlambda63}  provides the corresponding plot of $\bar{V}$ versus $\lambda_b$, with the red bullet points identifying the values $0.1,0.3,0.45,0.6,0.639$ of $\bar{V}$ on the curve.

\newpage


\section{Onset of instability\label{sec4}}


Consider the energy function $\omega^*(\lambda,D_{\mathrm{L}})$ given by \eqref{model-W-D} with the specialization \eqref{neo-Hookean} for the considered plane strain situation.  We denote by
 $\omega(\lambda ,E_{\mathrm{L}})$ the corresponding function with the independent variable $E_\mathrm{L}$, related to $\omega^*$ by
 $\omega^*(\lambda,D_{\mathrm{L}})=\omega(\lambda,E_{\mathrm{L}})+D_{\mathrm{L}}E_{\mathrm{L}}$.

From the analysis of Conroy Broderick et al. \cite{Conroy2020}, the \emph{second variation of the free energy density} of a homogeneously deformed plate can be written in terms of either $\omega^*$ or $\omega$ as
\begin{equation}
\omega^*_{\lambda\lambda}(\delta\lambda)^2+2\omega^*_{\lambda D_{\mathrm{L}}}\delta\lambda \delta D_{\mathrm{L}}+\omega^*_{D_{\mathrm{L}}D_{\mathrm{L}}}(\delta D_{\mathrm{L}} )^2 = \omega_{\lambda\lambda}(\delta\lambda)^2 -
\omega_{E_{\mathrm{L}}E_{\mathrm{L}}}(\delta E_{\mathrm{L}})^2,\label{sec-var-equi2}
 \end{equation}
 where $\delta\lambda$, $\delta D_{\mathrm{L}}$ and $\delta E_{\mathrm{L}}$, respectively, are variations in $\lambda$, $D_{\mathrm{L}}$ and $E_{\mathrm{L}}$, the corresponding subscripts representing partial derivatives.
For the inhomogeneous deformation considered here, this is also the local form of the second variation, and its
global counterpart, which requires integration over $X_1\in [0,A]$ and $X_2\in[-L/2,L/2]$, is, in terms of $\omega$,
 \begin{equation}
L \int_0^A[ \omega_{\lambda\lambda}(\delta\lambda)^2 -
\omega_{E_{\mathrm{L}}E_{\mathrm{L}}}(\delta E_{\mathrm{L}})^2]\mathrm{d}X_1=\frac{2AL}{\lambda_b^2-\lambda_a^2}\int_{\lambda_a}^{\lambda_b}[ \omega_{\lambda\lambda}(\delta\lambda)^2 -
\omega_{E_{\mathrm{L}}E_{\mathrm{L}}}(\delta E_{\mathrm{L}})^2]\lambda\mathrm{d}\lambda.
\label{integral-inhomog}
 \end{equation}

We focus initially on the local form of the second variation \eqref{sec-var-equi2}.  The Hessian matrix associated with the left-hand side of \eqref{sec-var-equi2}, denoted $\mathbf{H}^*$, is given by
\begin{equation}
\mathbf{H}^*=\left[\begin{array}{cc}
\omega^*_{\lambda\lambda}&\omega^*_{\lambda D_\mathrm{L}}\\
\omega^*_{\lambda D_\mathrm{L}}&\omega^*_{ D_\mathrm{L}D_\mathrm{L}}\end{array}\right].
\end{equation}
It has determinant
\begin{equation}
\det\mathbf{H}^*=\omega^*_{\lambda\lambda}\omega^*_{D_{\mathrm{L}}D_{\mathrm{L}}}-(\omega^*_{\lambda D_{\mathrm{L}}})^2,
\end{equation}
which can also be written $\omega_{\lambda\lambda}\omega^*_{D_{\mathrm{L}}D_{\mathrm{L}}}$, as shown in Dorfmann and Ogden \cite{DO2019} and Conroy Broderick et al. \cite{Conroy2020}.

For the quadratic form in \eqref{sec-var-equi2} to be strictly positive, $\mathbf{H}^*$ must be positive definite, so that
\begin{equation}
\omega^*_{\lambda\lambda}>0,\qquad \det\mathbf{H}^*\equiv \omega_{\lambda\lambda}\omega^*_{D_{\mathrm{L}}D_{\mathrm{L}}}>0.
\end{equation}
For the models
\begin{equation}
\omega^*(\lambda,D_{\mathrm{L}})=W(\lambda)+\frac{1}{2}\varepsilon^{-1}\lambda^{-2}D_{\mathrm{L}}^2,\qquad \omega(\lambda,E_{\mathrm{L}})=W(\lambda)-\frac{1}{2}\varepsilon\lambda^2E_{\mathrm{L}}^2,
\end{equation}
with $W$ given by \eqref{neo-Hookean}, we have
\begin{equation}
\omega^*_{D_{\mathrm{L}}D_{\mathrm{L}}}=\varepsilon^{-1}\lambda^{-2},\qquad \omega_{E_{\mathrm{L}}E_{\mathrm{L}}}=-\varepsilon\lambda^2.\label{omegastaromegamixed}
\end{equation}
Thus, for $\mathbf{H}^*$ to be positive definite, we require
\begin{equation}
\omega^*_{\lambda\lambda}>0,\qquad \omega_{\lambda\lambda}>0,
\end{equation}
and, in view of \eqref{omegastaromegamixed}$_2$, the global form of the second variation \eqref{integral-inhomog} is positive if $\omega_{\lambda\lambda}>0$ for each point of the domain of the integral.

On use of the connection $E_{\mathrm{L}}=\varepsilon^{-1}\lambda^{-2}D_{\mathrm{L}}$, the expressions \eqref{linear-law} and the formula \eqref{Vlast}, we find that $ \omega_{\lambda\lambda}$ is given by
\begin{multline}
\mu_a^{-1}(\lambda_b^2-\lambda_a^2)\lambda^4\omega_{\lambda\lambda} = (\lambda^4+3)(\lambda_b^2-\lambda_a^2+\alpha_\mu\lambda_a^2)-2\alpha_\mu(3\lambda^2-1)\lambda^4\notag\\[1ex]
- \frac{(\lambda_b^2-\lambda_a^2)^2(\lambda_b^2-\lambda_a^2-\alpha_\varepsilon\lambda_a^2+6\alpha_\varepsilon\lambda^2)[(1+\alpha_\varepsilon)\lambda_a^2-\lambda_b^2]^2\bar{V}^2}{[\lambda_b^2-\lambda_a^2+\alpha_\varepsilon(\lambda^2-\lambda_a^2)]^2\{\log[(1+\alpha_\varepsilon)\lambda_a^2/\lambda_b^2]\}^2},\label{wlamlam}
\end{multline}
and similarly, the expression for $\omega^*_{\lambda\lambda}$ is given by
\begin{eqnarray}
&&\mu_a^{-1}(\lambda_b^2-\lambda_a^2)\lambda^4\omega^*_{\lambda\lambda}=(\lambda^4+3)(\lambda_b^2-\lambda_a^2+\alpha_\mu\lambda_a^2)-2\alpha_\mu(3\lambda^2-1)\lambda^4\notag\\[1ex]
&&+\frac{(\lambda_b^2-\lambda_a^2)^2\{3[\lambda_b^2-\lambda_a^2+\alpha_\varepsilon(\lambda^2-\lambda_a^2)]^2+3[\lambda_b^2-\lambda_a^2+\alpha_\varepsilon(\lambda^2-\lambda_a^2)]\alpha_\varepsilon\lambda^2+4\alpha_\varepsilon^2\lambda^4\}
}{[\lambda_b^2-\lambda_a^2+\alpha_\varepsilon(\lambda^2-\lambda_a^2)]^3}\notag\\[1ex]
&&\times   \frac{[(1+\alpha_\varepsilon)\lambda_a^2-\lambda_b^2]^2\bar{V}^2}{\{\log[(1+\alpha_\varepsilon)\lambda_a^2/\lambda_b^2]\}^2}.\label{wstarlamlam}
\end{eqnarray}

When $\alpha_\mu=\alpha_\varepsilon=0$ and the deformation is homogeneous, we have
\begin{equation}
\mu^{-1}\lambda^4\omega^*_{\lambda\lambda}=\lambda^4+3+3\bar{D}_{\mathrm{L}}^2,\qquad \mu^{-1}\varepsilon\lambda^6\det\mathbf{H}^*=\lambda^4+3-\bar{D}_{\mathrm{L}}^2
=
\mu^{-1}\lambda^4\omega_{\lambda\lambda},
\end{equation}
where $\bar{D}_\mathrm{L}=D_\mathrm{L}/\sqrt{\mu_a\varepsilon_a}$, and we also note that $\bar{D}_\mathrm{L}=\lambda^2\bar{E}_\mathrm{L}$, where $\bar{E}_\mathrm{L}=E_\mathrm{L}\sqrt{\varepsilon_a/\mu_a}$.   Then, it is clear that $\omega^*_{\lambda\lambda}>0$, while $\det\mathbf{H}^*$ can become negative as $D_\mathrm{L}$ increases from zero, in which case stability according to the Hessian criterion is lost.  Since there is no applied load $\bar{D}_\mathrm{L}^2=\lambda^4-1$ and $\bar{E}_\mathrm{L}^2=1-\lambda^{-4}$ so that $\bar{D}_\mathrm{L}$ increases indefinitely with $\lambda$ while $\bar{E}_\mathrm{L}$ increases up to an upper limit $1$, i.e. a homogeneous plate in plane strain can only support a maximum non-dimensional voltage $\bar{V}=\bar{E}_\mathrm{L}=1$.  This is different from the equi-biaxial situation where $\bar{E}_\mathrm{L}^2$ exhibits a maximum and pull-in instability occurs; see, for example, \cite{Zhao2007,Su2018,DO2019}. 

Turning back to the inhomogeneous problem. 
\color{black}
In considering the inhomogeneity, for stability the integral expression in \eqref{integral-inhomog} must be positive for all possible choices of $\delta\lambda$ and $\delta E_{\mathrm{L}}$ with at least one of them non-zero \cite{Bertoldi2011}.   Since the second term in the integrand is non-negative and $\delta E_\mathrm{L}$ can be chosen to be zero, stability requires the integral
\begin{equation}\label{global_stability}
\int_{\lambda_a}^{\lambda_b}\lambda \omega_{\lambda\lambda}(\delta\lambda)^2 \mathrm{d}\lambda
\end{equation}
to be positive for all non-zero $\delta\lambda$.  If, however, $\omega_{\lambda\lambda}<0$  for part of the range of $\lambda$ then by choosing $\delta\lambda$ to be only non-zero for those $\lambda$ where  $\omega_{\lambda\lambda}<0$, the second variation would be negative, and we would have instability for such values of $\bar{V}$.  Thus, stability is guaranteed when $\omega_{\lambda\lambda}>0$ for every value of $\lambda\in[\lambda_a,\lambda_b]$.
\color{black} 

For the examples in Figures \ref{figure3} and \ref{figure44}, with positive values of $\alpha_\mu$ and $\alpha_\varepsilon$, we found \color{black} (not shown here) that $\omega_{\lambda\lambda}$ is negative for \emph{every} value of $\lambda\in[\lambda_a,\lambda_b]$ when $\bar{V}$ exceeds $0.1$, and for smaller values of $\bar{V}$ it is only positive for a very small range of values of $\lambda$ near $\lambda_a$.
Thus, such configurations are unstable according to both the local Hessian criterion and its global counterpart.  
We also found this instability for other positive values of $\alpha_\mu$ and $\alpha_\varepsilon$.

These interesting and quite surprising results therefore led us to consider if the situation was different for negative values of $\alpha_\mu$ and $\alpha_\varepsilon$.  It turns out, depending on the actual values of $\alpha_\mu$ and $\alpha_\varepsilon$, \color{black}  that the related configurations are stable for a range of values of $\bar{V}$. 

\begin{figure}[!h]
\centering
\includegraphics[width=\textwidth]{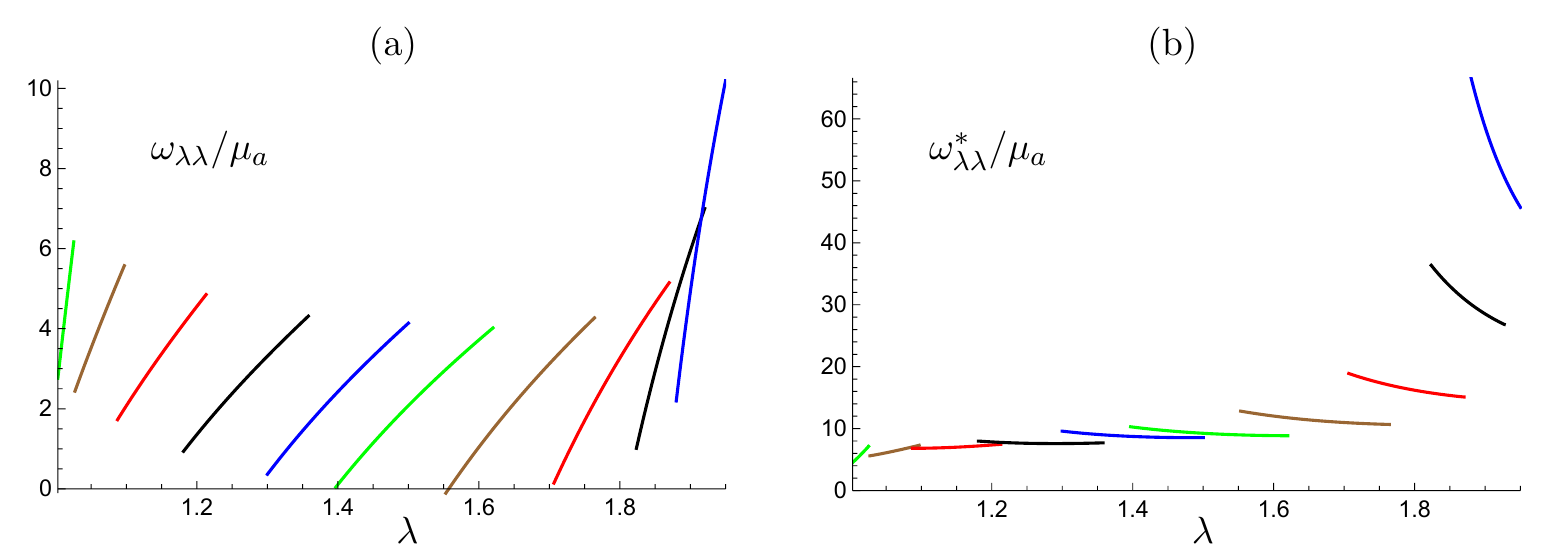}
\caption{Plots of  (a) $\omega_{\lambda\lambda}/\mu_a$ and (b)  $\omega^*_{\lambda\lambda}/\mu_a$ versus $\lambda$ for the following values of $\bar{V}$ reading from left to right: $0.1$, $0.3$, $0.5$, $0.65$, $0.75$, $0.8$, $0.85$, $0.88$, $0.895$, $0.9$, with  $\alpha_\varepsilon=-0.2$ and $\alpha_\mu=0.3$.   Note that range of values of $\lambda\in[\lambda_a,\lambda_b]$ is different in each case.
\label{wlambdalambdas}}
\end{figure}

 In Figures \ref{wlambdalambdas}(a) and (b), respectively, $\omega_{\lambda\lambda}$  and $\omega^*_{\lambda\lambda}$ are illustrated for a negative electric parameter $\alpha_\varepsilon=-0.2$ and positive elastic parameter $\alpha_\mu=0.3$, with several different values of $\bar{V}$ (refer to the bent shape in Figure \ref{rev2}(b) for $\bar{V}=0.3$, \color{black} for example\color{black}).    The range of values of $\lambda_a$ and $\lambda_b$ changes as the deformation increases with the voltage.    Both $\omega_{\lambda\lambda}$ and $\omega^*_{\lambda\lambda}$ remain positive as $\bar{V}$ increases except \color{black} that for a very small range of values of $\bar{V}$ near $0.85$ $\omega_{\lambda\lambda}$ is negative with maximum magnitude about $0.1$.  
In this case, the plate is unstable since, as has been shown in the discussion related to Eq. \eqref{global_stability}, stability requires that $\omega_{\lambda\lambda}$ be positive throughout the plate. \color{black}
With reference to the changing curvature shown in Figure \ref{minus2plus3} we mention that as $\bar{V}$ reaches a value between $0.903$ and $0.904$ the plate becomes straight and the curvature reverses for larger values of $\bar{V}$, but then the configuration becomes unstable (both  $\omega_{\lambda\lambda}$  and $\omega^*_{\lambda\lambda}$ become negative).  

\begin{figure}[!h]
\centering
\includegraphics[width=\textwidth]{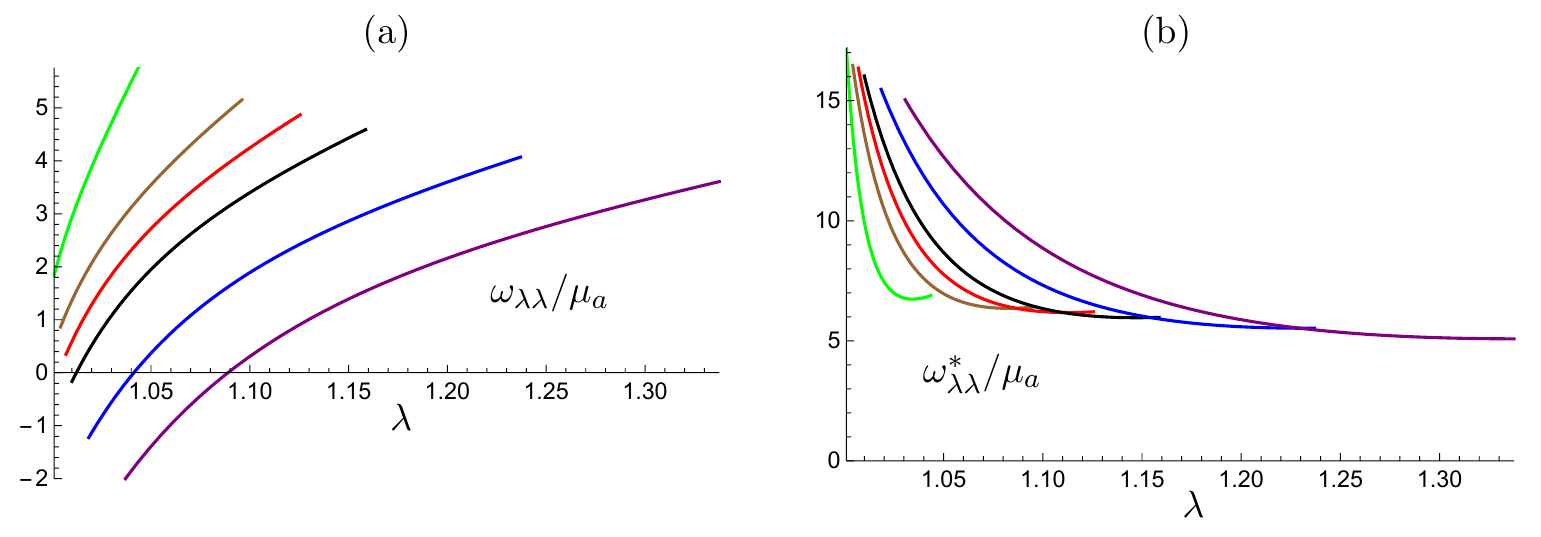}
\caption{Plots of  (a) $\omega_{\lambda\lambda}/\mu_a$ and (b)  $\omega^*_{\lambda\lambda}/\mu_a$ versus $\lambda$ for the following values of $\bar{V}$ reading from left to right: $0.1$, $0.2$, $0.25$, $0.3$, $0.4$, $0.5$, with  $\alpha_\varepsilon=-0.6$ and $\alpha_\mu=0.3$.   Note that range of values of $\lambda\in[\lambda_a,\lambda_b]$ is different in each case.
\label{wlambdalambdas2}}
\end{figure}

 In Figures \ref{wlambdalambdas2}(a) and (b), respectively, $\omega_{\lambda\lambda}$  and $\omega^*_{\lambda\lambda}$ are illustrated for $\alpha_\varepsilon=-0.6$ with $\alpha_\mu=0.3$ with several different values of $\bar{V}$ (refer to the bent shape in Figure \ref{minus6plus3}(b) for $\bar{V}=0.3$).  For $\bar{V}=0.1$, $0.2$, $0.25$, $\omega_{\lambda\lambda}$ is positive and hence, by \eqref{omegastaromegamixed}, the integral in \eqref{integral-inhomog} is positive, so the second variation of the free energy function is positive and the configuration is stable.    This is also the case for values of  $\bar{V}$ up to nearly $0.3$,  but $\omega_{\lambda\lambda}$ becomes negative for a small range of values of $\lambda$ when $\bar{V}=0.3$ and for a larger range with increasing $\bar{V}$, as illustrated by the blue and purple curves in Figure  \ref{wlambdalambdas2}(a).  On the other hand,
$\omega^*_{\lambda\lambda}$ is positive for all $\lambda$ and $\bar{V}$.

We emphasise that when $\omega_{\lambda\lambda}>0$ for $\lambda\in[\lambda_a,\lambda_b]$ the configuration in question is stable, and because  then $\omega^*_{\lambda\lambda} > \omega_{\lambda\lambda}$ there is no need to consider the sign of  $\omega^*_{\lambda\lambda}$ in assessing stability.  


\section{Concluding remarks\label{sec5}}


In summary, we have investigated the voltage-induced bending response of an FGDE plate, and formulated a method for analyzing the onset of its instability based on the loss of positive definiteness of the Hessian.   

We found that the `bendability' of the plate can be tuned by carefully selecting the grading properties, as illustrated here for  linear gradients, by choosing appropriate values of the inhomogeneity gradients $\alpha_\mu$ and $\alpha_\varepsilon$.  Depending on the signs of these parameters, a bent sector may be either (a) subject to compressive circumferential stresses at both its inner and outer surfaces, with two neutral axes, when both parameters are positive, or (b) subject to compressive circumferential stress on one surface and tensile stress on the other, with  a single neutral axis, when one of the parameters is negative.  On analysing the stability we found that configurations associated with (a) are always unstable, while configurations based on (b) are stable for certain combinations of $\alpha_\mu$ and $\alpha_\varepsilon$ or, for other combinations, stable provided the voltage is not too large.   

In contrast with the situation for a homogeneous plate subject to equi-biaxial deformations, where pull-in instability occurs, in the present plane strain problem there in no such pull-in instability.
In particular, even though the voltage-stretch curve exhibits a maximum, it cannot be associated with pull-in instability within the stable domain.

The analysis herein is based on a particular (neo-Hookean based) constitutive model and a simple form of the material grading properties.  It can be expected that the results will be somewhat different, at least quantitatively, for different model choices, but the framework presented here can accommodate more general models.

\color{black}
In this paper we have focused only on the possibility of instability arising from the loss of positive definiteness of the Hessian. 
This approach to analysis of stability tells only part of the story of stability since it does not allow for wrinkling types of instability in membranes \cite{Zurlo2017} or inhomogeneous wrinkling deformations \cite{Su2019c, DO2019}, which can be induced \cite{Mao2018} by the negative circumferential stress in the elastomer (Figures \ref{figure33}, \ref{figure45}, \ref{revstresses}). 
 \color{black}

Applications where gradient properties could be beneficial are in the design of sensors and actuators and complicated motions of soft robots, for example.


\section*{Acknowledgments}


This work was supported by a Government of Ireland Postdoctoral Fellowship from the Irish Research Council (No. GOIPD/2017/1208).




\begin{thebibliography}{99}


\bibitem{Pelrine2000}
R. Pelrine, R. Kornbluh, Q. Pei, J. Joseph,
High-speed electrically actuated elastomers with strain greater than 100\%,
Science 287 (2000), 836--839.

\bibitem{O'Halloran2008}
A. O'Halloran, F. O'Malley, P. McHugh,
A review on dielectric elastomer actuators, technology, applications, and challenges
J. Appl. Phys. 104 (2008), 071101.

\bibitem{Brochu2010}
P. Brochu, Q. Pei, Q,
Advances in dielectric elastomers for actuators and artificial muscles,
Macromol. Rapid Comm. 31 (2010), 10--36.

\bibitem{Alessandri2016}
I. Alessandri, J. Lombardi,
Enhanced Raman scattering with dielectrics,
Chem. Rev. 116 (2016), 14921--14981.

\bibitem{Nketia?Yawson2018}
B. Nketia-Yawson, Y. Noh,
Recent Progress on High-Capacitance Polymer Gate Dielectrics for Flexible Low-Voltage Transistors,
Adv. Funct. Mater. 28 (2018), 1802201.

\bibitem{Li2017}
T. Li, G. Li, Y. Liang, T. Cheng, J.  Dai, X. Yang, B. Liu, Z. Zeng, Z. Huang, Y. Luo, T. Xie, W. Yang,
Fast-moving soft electronic fish
Sci. Adv. 3 (2017), e1602045.

\bibitem{Li2018}
T. Li, Z. Zou, G. Mao, X. Yang, Y. Liang, C. Li, S. Qu, W. Yang,
Agile and resilient insect-scale robot,
Soft Robot 6 (2018), 133--141.

\bibitem{Wang2017}
L. He, J. Lou, J. Du, J. Wang,
Finite bending of a dielectric elastomer actuator and pre-stretch effects,
Int. J. Mech. Sci. 122 (2017), 120--128.

\bibitem{Su2019}
Y. Su, B. Wu, W. Chen, M. Destrade,
Pattern evolution in bending dielectric-elastomeric bilayers,
J. Mech. Phys. Solids 136 (2020), 103670.

\bibitem{Morimoto2015}
T. Morimoto, F. Ashida,
Temperature-responsive bending of a bilayer gel
Int. J. Solids Struct. 56 (2015), 20--28.

\bibitem{Rocca2010}
S. Roccabianca, M.Gei, D. Bigoni,
Plane strain bifurcations of elastic layered structures subject to finite bending:
Theory versus experiments,
IMA J. Appl. Math. 75 (2010), 525--548.

\color{black}
\bibitem{Knoppers2005}
G.E. Knoppers, J.W. Gunnink, J. Van Den Hout, W. Van Vliet,  
The reality of functionally graded material products, In Intelligent Production Machines and Systems: First I* PROMS Virtual Conference, Elsevier, Amsterdam (pp. 467-474) (2005).

\bibitem{Mahamood2012}
R.M. Mahamood, E.T. Akinlabi, M. Shukla, S. Pityana, Functionally graded material: an overview, Proceedings of the World Congress on Engineering 2012 Vol III, London, UK, (2012).

\bibitem{Wang2017}
X. Wang, M. Jiang, Z. Zhou, J. Gou, D. Hui, 
3D printing of polymer matrix composites: A review and prospective, 
Compos. Part B Eng. 110 (2017) 442-458.
\color{black}

\bibitem{Dorfmann2006}
A. Dorfmann, R.W. Ogden,
Nonlinear electroelastic deformations,
J. Elasticity 82 (2006), 99--127.

\bibitem{Dorfmann2014}
L. Dorfmann, R.W. Ogden, \emph{Nonlinear Theory of Electroelastic and Magnetoelastic Interactions}. Springer, 2014.

\bibitem{Zhao2007}
X. Zhao, Z. Suo,
Method to analyze electromechanical stability of dielectric elastomers,
Appl. Phys. Lett. 91 (2007), 061921.

\bibitem{Lu2012}
T. Lu, J. Huang, C. Jordi, G. Kovacs, R. Huang, D. Clarke, Z. Suo,
Dielectric elastomer actuators under equal-biaxial forces, uniaxial forces, and uniaxial constraint of stiff fibers,
Soft Matter, 8 (2012), 6167--6173.

\bibitem{Zhao2014}
X. Zhao Q. Wang,
Harnessing large deformation and instabilities of soft dielectrics: Theory, experiment, and application.
Appl. Phys. Rev. 1 (2014), 021304.

\bibitem{Zurlo2017}
G. Zurlo, M. Destrade, D. DeTommasi, G. Puglisi,
Catastrophic thinning of dielectric elastomers.
Phys. Rev. Lett. 118 (2017), 078001.

\bibitem{Su2018}
Y. Su, B. Wu, W. Chen, C. Lu,
Optimizing parameters to achieve giant deformation of an incompressible dielectric elastomeric plate,
Extreme Mech. Lett. 22 (2018), 60--68.

\bibitem{Su2019b}
Y. Su, W. Chen, M. Destrade,
Tuning the pull-in instability of soft dielectric elastomers through loading protocols,
Int. J. Non-Lin. Mech. 113 (2019), 62--66.

\bibitem{Rivlin1949}
R. S. Rivlin,
Large elastic deformations of isotropic materials V: the problem of flexure,
Proc. Roy. Soc. Lond. Ser. A 195 (1949), 463--473.

\bibitem{Su2019c}
Y. Su, B. Wu, W. Chen, M. Destrade,
Finite bending and pattern evolution of the associated instability for a dielectric elastomer slab,
Int. J Solids Struct. 158 (2019), 191--209.

\bibitem{Wu2017}
B. Wu, Y. Su, D. Liu, W. Chen, C. Zhang,
On propagation of axisymmetric waves in pressurized functionally graded elastomeric hollow cylinders,
J. Sound Vib. 421 (2018), 17--47.

\bibitem{Bayat2019}
M. Bayat, A. Kargar-Estahbanaty, M. Baghani,
A semi-analytical solution for finite bending of a functionally graded hydrogel strip,
Acta Mech. 230 (2019), 2625--2367.

\bibitem{DO2019}
L. Dorfmann, R.W. Ogden, Instabilities of soft dielectrics.  Phil. Trans. R. Soc. A 377 (2019), 20180077.

\bibitem{Conroy2020}
H. Conroy Broderick, M. Righi, M. Destrade, R.W. Ogden,
Stability analysis of charge-controlled soft dielectric plates. Int. J. Eng. Sci. 151 (2020), 103280.

\color{black}
\bibitem{Bertoldi2011}
K. Bertoldi, M. Gei,  
Instabilities in multilayered soft dielectrics. 
J. Mech. Phys. Solids 59(1) (2011), 18-42.

\bibitem{Mao2018}
G. Mao, Y. Xiang, X. Huang, W. Hong, T. Lu, S. Qu,  
Viscoelastic effect on the wrinkling of an inflated dielectric-elastomer balloon, J. Appl. Mech. 85(7) (2018).
\color{black}

\end{thebibliography}
\end{document}